\documentstyle[prl,aps]{revtex}
\begin{document}
\draft
\title{Conditional probabilities in quantum theory, and the tunneling time
 controversy}
\author{Aephraim M. Steinberg\footnote{Internet aephraim@physics.berkeley.edu;
Current address: Atomic
Physics Division, National Institute of Standards and Technology, Gaithersburg,
MD 20899}}
\address{Department of Physics, U.C. Berkeley, Berkeley, CA 94720}
\date{Preprint quant-ph/9502003; Submitted to Phys. Rev. A on July 19, 1994}
\maketitle
\begin{abstract}

It is argued that there is a sensible way to define conditional probabilities
in quantum mechanics, assuming only Bayes's theorem and standard
quantum theory.  These probabilities are equivalent to
the ``weak measurement'' predictions due to Aharonov {\it et al.}, and
hence describe the outcomes of real measurements made on subensembles.
In particular, this approach is used to address the question of the history
of a particle which has tunnelled across a barrier.  A {\it gedankenexperiment}
is presented to demonstrate the physically testable implications of the
results of these calculations, along with graphs of the time-evolution of
the conditional probability distribution for a tunneling particle and
for one undergoing allowed transmission.  Numerical results are also presented
for the effects of loss in a bandgap medium on transmission and on
reflection, as a function of the position of the lossy region; such loss
should provide a feasible, though indirect, test of the present conclusions.
It is argued that the effects of loss on the pulse {\it delay time} are related
to the imaginary value of the momentum of a tunneling particle, and it is
suggested that this might help explain a small
discrepancy in an earlier experiment.

\end{abstract}


\newcommand{\um}[1]{\"{#1}}
\renewcommand{\Im}{{\protect\rm Im}}
\renewcommand{\Re}{{\protect\rm Re}}
\newcommand{\ket}[1]{\mbox{$|#1\protect\rangle$}}
\newcommand{\bra}[1]{\mbox{$\protect\langle#1|$}}
\newcommand{\proj}[1]{\mbox{$\ket{#1}\bra{#1}$}}
\newcommand{\expect}[1]{\mbox{$\protect\langle #1 \protect\rangle$}}
\newcommand{\inner}[2]{\mbox{$\protect\langle #1 | #2 \protect\rangle$}}

The question of how much time a tunneling particle spends in the barrier region
has long been controversial, in part because it requires one to discuss not an
entire ensemble of identically prepared particles, but only the subset of
particles
which are later found to have tunnelled
\cite{Condon=1931,MacColl=1932,Eisenbud=1948,Wigner=1955,%
Hartman=1962,Stevens=1980,%
Buttiker=1982,Collins=1987,%
Hauge=1989,Buttiker=1990,Olkhovsky=1992,%
Leavens=1993BK,Landauer=1994}.  The absence of an unambiguous
prescription in quantum theory for dealing with such questions has compounded
the more technical problems, such as the negative kinetic energy of a
tunneling particle
 and the superluminal effects arising from the asymptotic
independence of the wave packet delay on barrier thickness
\cite{Steinberg=1993PRL,Enders=1993,Ranfagni=1993,Spielmann=1994,%
Nimtz=1994,Steinberg=1994COM,Steinberg=1995ANG}.

In essence, what is desired is a conditional probability distribution.  We know
how to ask what the probability is for a particle from a given ensemble to be
at point $x$ at time $t$.  Now we wish to ask for that same probability,
{\it conditioned} on the additional information that as $t \rightarrow \infty$,
the particle is found to have been entirely transmitted.  Conventional
wisdom holds that such distributions do not exist in quantum mechanics;
we can never have more information about a particle than its wave function.
Let us proceed nevertheless.

Consider the joint probability for two propositions $A$ and $B$, $P(A \& B)$.
This can be written as the probability of $B$, $P(B)$, multiplied by the
probability of $A$ {\it given $B$}, $P(A|B)$.  This is known as Bayes's
theorem, and will serve as our definition of conditional probabilites:
\begin{equation}
\label{Bayes}
P(A | B) \equiv \frac{P(A\&B)}{P(B)} \; .
\end{equation}
Now in quantum mechanics, the probability for a particle to be in a given
state can be expressed as the expectation value of the projector onto that
state;
\begin{eqnarray}
\label{projector}
P(A) & = & \expect{Proj(A)} = \expect{\proj{A}} \nonumber \\
& = & \inner{\psi}{A}\inner{A}{\psi} = | \inner{A}{\psi} |^2 \; .
\end{eqnarray}
For example, the probability that a particle is at point $x$ at time $t$ is
given by $|\inner{x}{\psi(t)}|^2 = |\psi(x,t)|^2$.
{}From a statistics perspective, we are looking for
the average value of a Boolean variable (the projector, whose eigenvalues
are 0 and 1), and interpreting the result as a frequency, or a probability.
If we wish to calculate the {\it joint} probability of $A$ and $B$, it
suffices to find the average value of the {\it product} of these two Boolean
variables:
\begin{equation}
\label{joint}
P(A\&B) = \expect{Proj(B) Proj(A)}
= \inner{\psi}{B}\inner{B}{A}\inner{A}{\psi}
\end{equation}
(where we have placed $B$ to the left of $A$ to indicate that it is to be
determined {\it after} $A$ in time).  This is where quantum mechanics diverges
 from standard probability theory, for the product of two Hermitian operators
need not be Hermitian-- in other words, P(A\&B) is in general complex.
While it has been noted before that there may be advantages to extending the
domain of probabilities beyond the region between 0 and 1, most physicists
are understandably reluctant to take such a step \cite{Muckenheim=1986,%
Youssef=1991,Scully=1994,Sokolovski=1991}.  It is not coincidental that
the situation here is analogous to the existence of imaginary momenta in
tunneling.  In that context, a quasiclassical approximation would suggest
that the traversal time $t = d/v = md/p$, becoming imaginary as well; certain
more careful calculations have also yielded complex values for the traversal
time.  It is natural to object that an imaginary number is unsatisfactory for
describing a time; as Landauer has put it, ``Has anyone seen a stop-watch
with complex numbers on its dial?''
In a similar vein, no one has rolled double-sixes an imaginary number of times.
However, quantum mechanics is at bottom not a
theory of probabilities, but rather one of wave functions.  A wave function,
even the wave function of the hand of a stopwatch, can be complex.  It is
only at the level of interpretation that these functions are turned into
probabilities.  In measurement theory, the justification for this is that an
``ideal'' measurement leads to complete decoherence between the various
possible outcomes, leaving only the (real) diagonal elements of the density
matrix to describe the system.  But quantum mechanically, the state of
a clock's hand may be $\exp [-(x-x_0)^2/4\sigma^2]$, where $x_0$ need
not be real.  It is in this sense that a quantum-mechanical clock may indeed
indicate a complex time.  At the level of observation, if we choose to measure
the
expectation value $\expect{x}$, we will find a real number: $\Re \;
x_0$.  If we wish to find $\Im \; x_0$, we must instead look at the
{\it momentum} of the clock hand, whose expectation value
is $\Im \; x_0 / 2\sigma^2$.

We therefore suspend our disbelief in complex probabilities, and
continue calculating the conditional probability distribution for the
position of a tunneling particle, $P(x,t | trans)$, where ``trans'' indicates
that the particle is found on the far side of the barrier (assumed to extend
from $-d/2$ to $+d/2$) at late times.  From
Eq. (\ref{Bayes}), we write this as
\begin{eqnarray}
\label{trans-prob}
& P(x,t | trans) = \frac{P(x,t\, \&\, trans)}{P(trans)} & \nonumber \\
& = \frac{\expect{Proj(trans)Proj(x,t)}}{\expect{Proj(trans)}} &\nonumber \\
& = \frac{\inner{\psi_i}{\psi_t}\bra{\psi_t}Proj(x,t)\ket{\psi_i}}
{\inner{\psi_i}{\psi_t}\inner{\psi_t}{\psi_i}} & \nonumber \\
& = \frac{\bra{\psi_t}Proj(x,t)\ket{\psi_i}}
{\inner{\psi_t}{\psi_i}} & \nonumber \\
& = \frac{1}{T} \psi^*_t(x,t) \psi_i(x,t) &  ,
\end{eqnarray}
where $\psi_i$ is the initial state in which the particle is prepared
({\it e.g.}, a packet incident from $x=-\infty$ at $t=-\infty$),
$\psi_t$ is the state of a transmitted particle ({\it i.e.}, at late times
simply $\psi_i$ projected onto the region on the far side of the
barrier), and $T \equiv \inner
{\psi_t}{\psi_i}$ is the transmission amplitude.  For a symmetric
barrier and symmetric initial conditions, $\psi_t$ is simply $\psi_i$
flipped about $x=0$ and about $t=0$, since it is defined to consist
of a packet heading towards $x=+\infty$ as $t \rightarrow +\infty$,
just as $\psi_i$ is defined to come entirely from $x=-\infty$ as
$t \rightarrow -\infty$.
In more practical terms, we can define $\psi_t(x,t) = T^* \psi_i(x,t) +
R^* \psi_i(-x,t)$ (where $R$ is the reflection amplitude), so that at
late times $\psi_t(x \gg 0,t \gg 0) \propto T^*T + R^*R = 1$ and $\psi_t
(x \ll 0,t \gg 0) \propto T^*R + R^*T = 0$.  Strictly speaking,
\begin{equation}
\psi_i(x,t) = \int_0^\infty \frac{dk}{\sqrt{2\pi}} \Phi(k) \psi_k(x,t).
\label{packets}
\end{equation}
$\Phi(k)$ is a bandwidth function, and the eigenstates
\begin{equation}
\psi_k(x,t)  =  \left\{
\begin{array}{ll}
\left[e^{ikx} + R(k) e^{-ikx}\right]
e^{-i\omega(k)t} & \mbox{for $x<-d/2$} \\
T(k) e^{ikx}e^{-i\omega(k)t} & \mbox{for $x>+d/2$}
\end{array}
\right.
\end{equation}
are orthogonal for different $k$.  Now,
\begin{eqnarray}
\psi_t(x,t) & \equiv & \int_0^\infty \frac{dk}{\sqrt{2\pi}} \Phi(k) \left[
T^*(k) \psi_k(x,t)
+ R^*(k) \psi_k(-x,t) \right] ; \; {\rm thus} \\
\psi_t(x<-\frac{d}{2},t) & = & \int_0^\infty \frac{dk}{\sqrt{2\pi}} \Phi(k)
\left[
T^*(k)e^{ikx}+T^*(k)R(k)e^{-ikx}+R^*(k)T(k)e^{-ikx}\right]e^{-i\omega(k)t}
 \nonumber \\
& = & \int_0^\infty \frac{dk}{\sqrt{2\pi}} \Phi(k) T^*(k)e^{ikx}
e^{-i\omega(k)t}
\nonumber \\
& = & \psi_i^*(-x,-t) \; \mbox{for $\Phi$ real}, \;  {\rm and} \\
\psi_t(x>\frac{d}{2},t) & = & \int_0^\infty \frac{dk}{\sqrt{2\pi}} \Phi(k)
\left[
T^*(k)T(k)e^{ikx}+R^*(k)e^{-ikx}+R^*(k)R(k)e^{ikx}\right]e^{-i\omega(k)t}
 \nonumber \\
& = & \int_0^\infty \frac{dk}{\sqrt{2\pi}} \Phi(k) \left[e^{ikx} + R^*(k)
e^{-ikx} \right]
e^{-i\omega(k)t} / \sqrt{2\pi} \nonumber \\
\label{phitransright}
& = & \psi_i^*(-x,-t) \, \; \mbox{for $\Phi$ real.}
\end{eqnarray}
(The assumption of $\Phi$'s reality assures the time-symmetry of the
incident state, yielding the parity relationship between the various
states.)
So long as $\Phi(k)$ is restricted to a bandwidth over which the transmission
and reflection coefficients vary negligibly, we may approximate the latter
as constants, and Eq. \ref{phitransright} does indeed correspond to the
projection of $\psi_i$ onto the far side of the barrier at late times
(for a more careful discussion of the effects of finite bandwidth
on the closely related Larmor clock, see \cite{Falck=1988}).  It is also
simple to confirm that
\begin{equation}
\inner{\psi_t}{\psi_i} = \int_0^\infty dk |\Phi(k)|^2 T(k) \rightarrow T \; .
\end{equation}


To find a ``conditional expectation value'' for an observable $R$ in
general, for a particle prepared in state $i$ and later found to be
in state $f$, we sum over $R$'s eigenvalue spectrum:
\begin{equation}
\label{weak}
\expect{R}_{fi} \equiv \sum_j P(R_j|f) r_j = \frac{\bra{f}R\ket{i}}
{\inner{f}{i}} \; ,
\end{equation}
where $R_j$ and $r_j$ are the eigenstates and corresponding
eigenvalues of $R$.  The question to be answered is whether this
expression, which follows from blindly applying probability rules
to quantum mechanical expressions which need not be real-valued, has
any physical meaning.  Indeed, Aharonov {\it et al.} have arrived
at precisely this expression on entirely physical grounds
\cite{Aharonov=1990,Aharonov=1988}.  They
reexamined von Neumann's theory of measurement, considering
a new limit which they termed ``weak measurement.'' In this regime,
the measuring device is prepared in a state which will disturb the
system to be measured as little as possible.  As a result of the
uncertainty principle, each individual measurement of this type is
extremely imprecise.  Aharonov {\it et al.} asked nevertheless what
the mean result would be for a large number of such measurements,
if one only examined the measuring device on those occasions on
which the system being studied was found to be in the
desired final state $f$ ({\it after} the period
during which the measurement interaction was on).

They considered a measuring apparatus or ``pointer'' whose position and
conjugate momentum we shall term $Q$ and $P$.  A measurement of
a given observable $R$ results from
a time-dependent interaction
\begin{equation}
\label{vN}
{\cal H}_{int} = g(t) P \cdot R \; ;
\end{equation}
 since $P$ is the
generator of translations for the pointer, the mean position of the pointer
after the interaction will have shifted by an amount proportional to the
expectation value $\expect{R}$.  In an ideal measurement, the relative
shifts corresponding to different eigenvalues of $R$ are large compared
with the initial uncertainty in the pointer's position, and the resulting lack
of overlap between the final states leads to the effective decoherence
(or irreversible ``collapse'') between different eigenstates of $R$.  In
\cite{Aharonov=1990,Aharonov=1988}, this approach is modified
in that the initial position of the pointer has a large uncertainty, so
that the overlap between the pointer states which become entangled with
the state of the particle
remains close to unity, and hence that the
measurement does not constitute a collapse.
Seen another way, this means that the pointer
momentum $P$ may be very well-defined, and therefore need not impart
an uncertain ``kick'' to the particle; the measurement is ``weak'' in that it
disturbs the state of the particle as little as possible between the state
preparation and the post-selection.  If the initial state
of the pointer is $\exp [-Q^2 / 4\sigma^2]$, then after the measurement it
will, for suitably normalized $g(t)$, be in the state
$\exp [-(Q-\expect{R}_{fi})^2 / 4\sigma^2]$.  As discussed
earlier, the real
part of $\expect{R}_{fi}$ corresponds to the mean shift in the pointer
position, while the imaginary part constitutes a shift in the pointer {\it
momentum}.
This latter effect is a reflection of the
back-action of a measurement on the particle.  It
obviously does have physical significance, but since it does not
correspond to a spatial translation of the pointer, should {\it not}
 be thought of
as part of the measurement outcome.  Furthermore, unlike the spatial
translation $\Delta Q = \Re \expect{R}_{fi}$, this effect is sensitive to the
initial state of the pointer: $\Delta P = \Im \expect{R}_{fi} / 2\sigma^2$.
As $\sigma$ becomes large, the measurement becomes very weak, and
the momentum shift of the pointer (like the back-action on the particle)
vanishes, while the spatial shift remains constant.

It is already quite remarkable that the same expressions for these
conditional probabilities or ``weak values'' arise from general probability
arguments as well as from careful consideration of measurement
interactions.  These expressions have many other attractive properties
which make it tempting to consider them ``elements of reality.'' In
many respects, they behave in a more intuitive fashion
than do wavefunctions themselves.  For
example, in \cite{Aharonov=1990} it is pointed
out that if a ``weak'' measurement
of an operator $R+S$ is made on a particle after it is prepared in an
eigenstate of $R$ with eigenvalue $r$ and before it is detected in an
eigenstate of $S$ with eigenvalue $s$, the result will be simply
$r+s$.  This holds whether or not $R$ and $S$ commute, and
even if $r+s$ is outside the eigenvalue spectrum
of $R+S$; hence such a simple rule could not be obeyed by standard
quantum measurements (ones which are precise, or ``strong,''
 enough to disturb the
time-evolution between $r$ and $s$).  More generally, weak values are
noncontextual and additive;
 $\expect{R+S}_{fi}=\expect{R}_{fi}+\expect{S}_{fi}$.
When averaged over an orthonormal set of final states, they reproduce
the usual expectation value, since $P(A)=\sum_f P(f) P(A|f)$.  These
conditional probabilities are also easily shown to obey a chain
rule, $P(A\&B|f) = P(B|f) P(A|B)$.  In sum, there are many reasons
to ascribe a certain level of reality to these conditional probability
distributions.

In \cite{Steinberg=1994WK},
it was argued that one could use this formalism to calculate
the ``dwell time'' for transmitted or reflected particles individually.  In
the time-independent case treated in that paper, the dwell time is defined
as the number density integrated over the barrier region (the expectation
value of the projector onto the barrier region, $\Theta_B \equiv
\Theta(x+d/2) - \Theta(x-d/2)$ for a barrier extending between $\pm d/2$)
divided by the incident flux.  It was shown for a rectangular barrier that
these conditional dwell times were equal to the usual dwell time calculated
for the ensemble as a whole, plus an imaginary part corresponding to
the back-action of a measuring device on the tunneling particle.
For the transmitted particles, however,
the real part of this dwell
involved equal contributions from regions near either edge of
the barrier, while for reflected particles the dominant contribution came
from the region near the entrance face only.  Here I would like to
expand on the utility of this approach to tunneling times, by presenting
complete probability distributions for the position of a particle at a given
time, conditioned upon reflection or transmission.  These can be
related back to the conditional dwell times by
\begin{equation}
\label{taut}
\tau_f = \int_{-\infty}^{+\infty} dt \int_{-d/2}^{+d/2} dx P(x,t | f) \; ,
\end{equation}
where $f$ indicates the final state under consideration and $\pm d/2$ are
the edges of the barrier.

To understand the meaning of these conditional probability distributions,
let us first consider an explicit {\it gedankenexperiment}.
In \cite{Steinberg=1994WK}, it was seen that the Larmor times
\cite{Baz=1967,Rybachenko=1967,Buttiker=1983}
are simply one instance of a ``weak measurement'' of the dwell
time.  The two components of the Larmor time turn out (as in
the closely related analysis by Sokolovski {\it et al.}
\cite{Sokolovski=1991,Sokolovski=1987,%
Sokolovski=1990,Sokolovski=1993,Sokolovski=1994}; see also
\cite{Fertig=1990,Fertig=1993,Muga=1992,Brouard=1994}) to be the real and
imaginary parts of the conditional dwell time.  By considering
the effects of preparing the measuring device (in that case, the spin
of the tunneling particle itself) in a state with great ``pointer position''
uncertainty, it was found that only the real part (representing
the mean shift in pointer position, and neglecting the pointer
momentum) had physical
significance independent of the details of the measurement.
However, the fact that both the ``position'' and the
conjugate ``momentum'' of that pointer are simply spin angles
makes the distinction a subtle one.  In addition, that experiment
fails to show how the conditional probability distributions for
transmitted and reflected particles differ, measuring only the
integrated dwell time, which is the same for both subensembles.
Instead of considering spin, let us imagine a scenario as in
Fig. 1 \cite{Steinberg=1994NYAS}.  A heavy charged particle
such as a proton is tunneling in one
dimension.  It passes through a series of parallel conducting plates with
small holes.  The plates may be held at a large positive voltage in
order to form a tunnel barrier for the proton.  (Due to the attraction
between the proton and its image in each plate, it also experiences
a periodic effective potential; in principle, if the plates were separated
by one-half the proton's de Broglie wavelength, they would therefore
form a ``bandgap'' for the proton even if no external potential were
applied.)  We will be concerned with the ``opaque'' limit, {\it i.e.},
the case where the transmission probability is small because the
evanescent decay constant $\kappa =\sqrt{k_0^2-k^2}$ (where
$k$ is the incident wavevector and $\hbar^2 k_0^2 /2m$ is the
height of the barrier) is much greater than the reciprocal of the
barrier width.  Between each pair of plates, but far from the proton's
trajectory, is an electron constrained to move parallel to the plates.
Due to the shielding of the proton's field by the plates, each electron
only feels a significant Coulomb force
while the proton is between the same pair of plates as that electron.
Thus each electron serves as a test particle; if it begins at rest, its
final momentum serves as a record of how long the proton spent
in the region from which it is not shielded.
Note that the electron {\it momentum} is thus the ``pointer position,'' and
the conjugate ``pointer momentum'' is the physical {\it position} of the
electron.  This follows from the form of the Coulomb interaction, which
in the presence of the conducting plates can be written
\begin{equation}
\label{Coulomb}
{\cal H}_{int} = -e^2 g(x_p) / y_e \; ,
\end{equation}
which is approximately linear in electron transverse position $y_e$ (so
long as $\Delta y_e \ll \expect{y_e}$)
and is proportional to a function of the proton's longitudinal position,
$g(x_p)$, which is close to zero except in
the region between the two plates.
Let us fire protons through this apparatus one by one, and first examine
the electrons only on those occasions when a proton is reflected.
(The electrons are all ``reset'' between shots.)  The momentum shift
of the electron is now proportional to the ``weak value'' of the time
spent by the proton in its region of sensitivity.  In other words, it
measures the time integral of the conditional probability that the electron
was in its region, given that the proton was to be reflected:
\begin{equation}
\Delta p_e \propto \int dt \int dx \; g(x) P(x,t|refl) \; ,
\end{equation}
where $g(x)$ describes the force on the electron for a proton at position $x$,
and is reasonably well confined to the region between the pair of
plates surrounding the electron in question.
The arrows in the figure indicate the final momentum of the electron.
What we find is perhaps not unexpected.  The proton's wave function
decays exponentially inside the barrier, and so does $P(x|refl)$; only the
electrons closest to the entrance face accumulate a significant momentum
kick.  Since nearly all the particles are reflected, this conditional
probability
is nearly the same as the unconditioned probability $|\psi(x)|^2$, and thus
has a negligible imaginary part; none of the electrons undergoes a significant
{\it position} shift aside from the time-dependent one due to their
final momenta.

But now what if we consider only events where the proton is transmitted?
Here we find that $P(x|trans)$ is essentially an even function of $x$,
as can be seen by examining Eq. (\ref{trans-prob}) and recalling that
$\psi_i$ and $\psi_t$ are related by a parity flip (along with a
time-reversal).
Using our more practical definitions of $\psi_t$ and $\psi_r$, we have
\begin{eqnarray}
\label{probs2}
P(x,t|trans) = \frac{1}{T} \psi^*_t(x,t) \psi_i(x,t)
= |\psi_i(x,t)|^2 + \frac{R}{T} \psi^*_i(-x,t)\psi_i(x,t) \nonumber \\
P(x,t|refl) = \frac{1}{R} \psi^*_r(x,t) \psi_i(x,t)
= |\psi_i(x,t)|^2 + \frac{T}{R} \psi^*_i(-x,t)\psi_i(x,t) \; .
\end{eqnarray}
One can see clearly from this that $|T|^2P(x,t|trans) + |R|^2P(x,t|refl)
= |\psi_i(x,t)|^2$ as expected, which leads directly to the well-known
relation between transmission, reflection, and (full-ensemble) dwell
times $|T|^2\tau_T + |R|^2\tau_R = \tau_d$.  One can also see that for
$|R| \gg |T|$, as in the opaque limit, $P(x,t|refl)$ is essentially equal
to the absolute square of the incident wave function (decaying exponentially
into the barrier), while $P(x,t|trans)$ is dominated by a term which
is an even function of $x$.  After more careful consideration, one
also notes that since $R$ and $T$ are $90^\circ$ out of phase
\cite{Falck=1988,Ou=1989AJP,Steinberg=19941D2D} and since
$\psi_i$ is dominated in the barrier region by real exponential decay,
$P(x,t|trans)$ is mostly imaginary.  Only near the two extremes of
the barrier, where the differing phases of the evanescent and anti-%
evanescent waves become important, does the real part become
significant.  As shown in \cite{Steinberg=1994WK},
\begin{eqnarray}
\label{times}
\tau_T & = & \frac{m}{\hbar k} \, \frac{1}{T} \left[
(B^2  + C^2) d + (BC+CB) \frac{\sinh \kappa d}{\kappa} \right]
 \nonumber \\
\tau_R & = & \frac{m}{\hbar k} \, \frac{1}{R} \left[
(BC  + CB) d + (B^2+ C^2) \frac{\sinh \kappa d}{\kappa} \right]
\nonumber \\
\tau_d & = & \frac{m}{\hbar k} \,  \left[
(B^*C  + C^*B )d + \left(|B|^2+
|C|^2\right) \frac{\sinh \kappa d} {\kappa} \right]
 \; ,
\end{eqnarray}
where $B$ and $C$ are the coefficients of the evanescent and anti-%
evanescent waves, respectively, and satisfy $|B/C| =exp[\kappa d] \gg 1$.
In Fig. 1b, the effect of this is seen.  The overall momentum transferred
to the set of electrons is the same as in the case of a reflected proton,
but it is now split evenly between the electrons within an exponential
decay length of {\it either} edge of the barrier; when a particle is
transmitted,
it spends as much time by the exit face of the barrier as by the entrance
face.  The electrons in the center of the barrier still exhibit essentially
no momentum shift!  It is as though the proton simply ``hopped'' from
one edge to the other, spending negligible time in the barrier.  This is
related to the well-known fact \cite{Hauge=1989}
 that the wave packet delay time in
opaque tunneling is independent of the barrier thickness.  However,
we must recall that the {\it imaginary} part of $P(x,t|trans)$ is significant
over the entire barrier.  This manifests itself as a shift in the mean
{\it position} of all the electrons (which we have been describing as the
pointer ``momentum,'' and which is never affected by ``ideal'' quantum
measurements).
This is simple to understand; the electrons have some uncertainty in their
position to begin with.  Due to the attractive Coulomb interaction, the
closer a given electron is to the proton, the smaller the potential barrier the
proton has to traverse.  Thus by selecting protons which succeeded in
tunneling, we are post-selecting states where the electrons were nearby to
begin with.  The constancy of this back-action across the length of the
barrier reflects the fact that within the WKB approximation, the transmission
is $exp[-\int \kappa(x) dx]$, {\it i.e.}, equally sensitive to a change in the
potential at any point in the barrier.
Unlike the momentum kick-- the measurement outcome
itself-- this effect is entirely dependent on the initial uncertainty in the
electron position.  If the initial states of the electron are very well
localized
in space, they are hardly shifted at all by this effect.  On the other hand,
the uncertainty principle then requires their initial momenta to be poorly
defined, ``weakening'' the resolution of the measurement.  Over many
trials, however, the mean value of the momentum shift will be unaffected
by the choice of initial uncertainty.

As discussed in \cite{Steinberg=1994WK},
this pointer position shift corresponds
to the in-plane portion of the Larmor time, while the pointer momentum shift
corresponds to minus the out-of-plane portion
\cite{Buttiker=1983,Sokolovski=1987}.  Both time scales are
 meaningful,
but their meanings are distinct.  (Furthermore, the suggestion that one should
pay attention to $|\tau_T|$ rather than to $\tau_T$ itself
seems odd, if only because unlike the real and
imaginary parts individually, this fails to satisfy the stipulation that the
dwell
in a large region be equal to the sum of the dwell in a set of smaller regions
which make it up.)  The real part of $\tau_T$ indicates the magnitude
of the effect our tunneling particle would have on test particles.  It also
describes the amount of absorption or gain the tunneling particle(s) would
suffer if tunneling through an active medium.  For $k \stackrel{>}{_\sim}
\kappa$, this time also approaches the group delay time for the peak
of the tunneling packet.  On the other hand, the imaginary part indicates
the magnitude of the {\it back-action} on the tunneling particle {\it due
to the measurement.}  This is the timescale which emerges from consideration
of the effects of an oscillating barrier on a tunneling particle, for example
\cite{Buttiker=1982,Martin=1993}.  Since unlike the real part, it
grows proportionally
with barrier thickness, it dominates in the opaque limit, where it reduces
to $md/\hbar\kappa$.  (It is interesting to note that to good approximation,
the ``weak value'' thus follows what one might expect in the WKB limit,
and yields $\expect{\tau_T}=md/\expect{p}$, even when $\expect{p}$ is
imaginary; this is another example where the behavior of weak values
obeys simple rules even in regimes where we expect these rules to break
down; see also \cite{Aharonov=1993NKE}.)
Importantly, the back-action due to this imaginary part
 depends strongly on the
initial state of the measuring device; in this sense, it is not a
characteristic
of the tunneling particle itself.  Of course, in real devices, there will be
an interplay between these two timescales.  The tunneling particle will
affect nearby particles, which may in turn modify the tunneling
characteristics.
The former effect, however, can now be seen to be limited to the extremities
of the barrier region, while the latter can be arbitrarily small if the nearby
particles are arranged in such a way that they do not create a large
uncertainty
in the potential seen by the tunneling particle.

In Fig's 2 and 3, we can observe the time-evolution of the conditional
probability distributions.  At early times, the
distribution is real, and mimics the initial wave packet; at late times, it
is also real, and mimics the final wave packet.  This seems to be in contrast
 to the claim that in order to reconcile superluminal peak propagation with
causality, we must consider all the transmitted particles to have originated
near the leading edge of the incident wave packet \cite{Leavens=1993BK,%
Deutch=1993,Steinberg=1994HI}.  Figure
2 presents the real and imaginary parts of $P(x|trans)$ for a wavepacket
incident from the left on a barrier which extends from $x=-5$ to $x=+5$.
Also shown is the full-ensemble distribution $P(x) \equiv |\psi(x)|^2$.
The wave packets are constructed according to Eq. \ref{packets}, with
a real Gaussian bandwidth function whose parameters are given in the
figure captions.
For the parameters chosen, the transmissivity is about $5.5 \cdot 10^{-6}$,
so $P(x|refl)$ would be essentially indistinguishable from $P(x)$.  As
advertised, the real part of $P(x|trans)$
is exponentially suppressed near the center of the barrier,
while the imaginary part becomes large and essentially constant across the
barrier region during the tunneling event.  Oscillations in both parts are
seen to either side of the barrier, due to the self-interference of the wave
packet near the potential step.  As discussed in \cite{Steinberg=1994WK},
these extra-barrier oscillations average out to zero when integrated over
space; the total time spent by a particle in a long region $L$ to either side
of the barrier is simply $mL/\hbar k$.  The regions of negative
conditional probability have a clear physical meaning; in the
experiment of Figure 1, for example, they predict that a test electron
would experience a {\it repulsive} momentum transfer
rather than an attractive one
due to the proton's Coulomb field.  Similar predictions apply to any
other von Neumann-style measurement one might contemplate.  For this
reason, although we sympathize with those who consider negative times
``unphysical,'' we do not see a better definition than that provided
by measurement outcomes; to deny this definition, one would either have
to give up the expression for the Coulomb force or the idea that momentum
transfer is the force integrated over the interaction time.
{}From the figures, we can also see that as the wavepacket
overcomes the barrier, the peaks in $P(x|trans)$ decrease in magnitude
on the left side of the barrier, and grow on the right side, without ever
traversing the center.  This is what was meant after Eq.
\ref{times}  by the particle ``hopping'' across the barrier.  It is
in a sense ``nonlocality of a single particle.''  While it is well known
that despite  the nonlocality inherent in quantum mechanics, no
expectation values can ever depend on choices made at spacelike separated
points (and thus that nonlocality does not violate Einstein causality),
this shows that a single particle can affect expectation values of
(weak) measuring devices at two spacelike separated positions (see
related discussions of the ``reality of the wave function''
\cite{Aharonov=1993AV,Aharonov=1993AAV,Unruh=1994}).

It is worth asking whether the ``duration'' of the tunneling process is best
defined by Eq. (\ref{taut}), or rather by the length of time over which
$P(x \in [-d/2,d/2] ,t| trans)$ remains close to its maximum value.  As
can be seen from the figures and from Eq. (\ref{trans-prob}), the latter
quantity is
simply the length of time during which the wave function is significant
at the location of the barrier.  Since in tunneling, the group delay time
for traversal is small compared with the temporal width of the freely
propagating wave packet width (the contrary
would imply $\tau_g \approx 2m/\hbar k\kappa
\stackrel{>}{_\sim} (1/\Delta k)/(\hbar k/m)$,
or $\Delta k \stackrel{>}{_\sim}\kappa$, leading to $(k + \Delta k)^2
> k^2 + \Delta k^2 > k^2 + \kappa^2 = k_0^2$, {\it i.e.}, a significant
portion of the wave packet having enough energy to traverse the
barrier without tunneling), this time scale is dominated by the duration
of the wave packet, approximately $m/\hbar k \Delta k$.  It should be
borne in mind that the same would be true for free propagation.  If a
wave packet of length $L$ traverses an empty region of width
$d \ll L$, the length of time during which a change in the potential in the
region could significantly affect the transmission probability is $L/v$;
nonetheless, for most intents and purposes we would think of an
individual particle as spending only $d/v$ in that region.  Obviously,
the relevant timescale will depend on the precise experimental question
one wishes to address.

Figure 3 shows the case for a particle which is incident with enough energy
to traverse the barrier.  The reflection probability is about 3.5\%, and
both $P(x|trans)$ and $P(x|refl)$ are displayed, along with $P(x)$.  The
latter is generally obscured by $P(x|trans)$, but in those regions where it
can be seen, it is clearly the weighted average of the reflection and
transmisison
distributions, as expected.  In this figure, we find the transmitted portion
traversing the barrier relatively smoothly, albeit with some oscillations.
It has a very small imaginary part (as follows for $|T/R| \approx 5$).
The reflected portion, on the other hand, undergoes violent oscillations in
both its real and imaginary parts.  At late times, these die away, and a
real wave packet propagating back to the left is all that remains.

One might think that the real and imaginary parts of this weak value,
one part describing the effect of
a particle on a measuring device and the other describing the back-action
of the measuring device on the particle, should be equal in the ``classical''
limit, that is, for allowed transmission.  Inspecting the figures, however,
one sees that in the region of allowed propagation, the imaginary part
tends to zero.  This can be understood as follows: an infinitesimal
perturbation
in the potential leads to a reflection amplitude proportional to the
perturbation.
If prior to the perturbation, the reflectivity was zero, the reflection
probability
is quadratic in the perturbation and hence vanishes in the limit of a gentle
measurement.  If on the other hand, there is a non-zero reflectivity prior
to the perturbation, then the reflection probability $|r_0 + \delta r|^2$ grows
linearly with the perturbation, indicating a finite back-action.

As discussed in \cite{Steinberg=1994WK}, another interesting feature of
these weak values is that $P(x>d/2 | refl)$ need not vanish.  This
implies that a particle incident from the left and ultimately reflected
does spend some time to the right of the barrier, at least within one
wave packet width of the exit face.  Although this can be understood
as an effect due to the possibility of coherent reflection off the
measuring device itself, the fact that the calculated time is independent
of the type or strength of the measurement interaction may be taken
to ascribe a certain level of reality to this time regardless of how or
even whether it is observed.  The argument is essentially the same one
made in favor of negative times.  It should be noted that as the strength
of the measurement interaction is lowered, the probability of a particle
being reflected {\it by} the measuring device falls as the square of
the potential, while the ``pointer position'' shift is linear in the
interaction strength.

Although it is always possible to argue about definitions of words
like ``interaction time,'' the advantage of the weak measurement approach
is that it offers a simple and intuitive formalism to treat a broad
variety of experimental predictions in a unified manner.  While some
will not want to call a negative or a complex number a ``time,'' the
words are relatively unimportant; the features shown in Figures 3 and 4
are in principle experimentally testable.
In particular, the result shown schematically in Fig. 1, that transmitted
particles have equal effects near both sides of a barrier, while reflected
particles affect only the region around the entrance face, is to my
knowledge a new prediction.  The {\it gedankenexperiment} of that
figure is of course unrealistic, but it is conceivable that in the future
an analogous
experiment could be performed.   For instance, one might consider
Rydberg atoms travelling through
two or three successive micromasers and leaving partial information in
the stored fields.

At the present time, there is an indirect test which should be feasible and
in fact rather straightforward.  As demonstrated in \cite{Steinberg=1993PRL,%
Steinberg=1995ANG},
a multilayer dielectric mirror possesses a photonic bandgap
\cite{Yablonovitch=1993}
and may be used as a tunnel barrier.  These mirrors are made of alternating
high- and low-index quarter-wave layers.  Typically, each dielectric layer has
very low loss, and the low transmissivity is due to reflection rather than to
absorption.  If one layer is now doped with an absorbing material, in general
both the transmission and the reflection should decrease; the amount of the
decrease can be thought of as an indication of the time spent by the
transmitted
or reflected photons in the layer in question.  Figure 4 contains numerical
results for an 11-layer mirror.  The ratio of the transmission (and reflection)
in the presence of one absorbing layer to the transmission (and reflection) in
the absence of absorption was calculated.  The logarithm of this ratio
(measured in units of the single-layer attenuation, so that in semiclassical
terms what is being plotted might be thought of as the number of passes
through the layer)
 is plotted versus the position of the lossy layer.
Fig. 4(a) is for light incident in the center of
the bandgap, with transmission of about 1.2\% (in the absence of any absorber).
The similarity of these curves to those of Fig. 2 should be evident.  In Fig.
4(b),
the light is incident outside the bandgap, near the first resonant transmission
point
(where the entire mirror may be thought of as a single Kronig-Penny crystal,
and multiple reflections between the opposite edges of the mirror interfere
constructively, leading to near-100\% transmission).  In this regime,
the mirror is essentially
a low-Q Fabry-Perot operating in its fundamental mode, and the transmission
is most sensitive to an absorber near the center of the structure, where the
mode has its maximum.  The reflection is close to zero in the absence of an
absorber, so the addition of absorption can only lead to effective gain in the
reflection channel; this is the meaning of the negative value of the curve.  In
Fig. 4(c), the light is still outside the bandgap, but now at a transmission
minimum of about 66\%.  One can again observe the mode structure within
the barrier, and complementary oscillations for the transmitted and reflected
parts, reminiscent of Fig. 3.

The effect of absorption offers another way of understanding the meaning
of complex times.  If each quarter-wave layer has an amplitude transmissivity
of
$1-\alpha$, we may think of this as an amplitude of $\exp[-2\alpha\omega
 t/\pi]$ to survive a time $t$ spent in the barrier.  In the tunneling regime,
the time
becomes predominantly imaginary; $\tau_t \sim -i\tau_{BL}$, where
 $\tau_{BL}$ is the B\"uttiker-Landauer time ($md/\hbar \kappa$ for a massive
particle).  Thus the attenuation factor becomes a phase shift rather than
absorption.  This is connected to the weak measurement idea in that if
absorption
is used as a clock, the ``pointer'' is essentially the photon number.  The
conjugate
``momentum'' is hence the optical phase.  Yet another way of understanding this
is to recall that the imaginary part of the ``conditional'' dwell time, like
the
out-of-plane Larmor rotation, can be related to an energy-derivative
 of the {\it magnitude}
of the transmission amplitude,
while the real part, like the usual dwell time, the group delay, and the
in-plane Larmor rotation, can be related to an energy-derivative of the {\it
phase}
of the transmission amplitude  \cite{Buttiker=1983,Steinberg=1994WK}.
Since absorption (or gain, which will behave in precisely the same way)
can be expressed as an imaginary contribution to an effective Hamiltonian,
we see by analytic continuation that the real part of the conditional dwell
time describes the amount of attenuation caused by an absorber,
and the imaginary part describes the absorber's effect on the phase.  An
interesting corollary to this is that loss within the tunnel barrier may
increase
the physical delay (cf.
\cite{Ranfagni=1990,Mugnai=1992,Raciti=1994,Nimtz=1994DIS})
by introducing a
frequency-dependent
phase shift, in principle without a large impact
on the transmission and reflection probabilities themselves.  Indeed, in
 \cite{Steinberg=1993PRL}, the measured delay time exceeded the group
delay prediction by about 0.4 fs, but this was at the borderline of statistical
significance.  More recently, we have found this discrepancy
to be statistically significant and to persist
for two different barriers of identical design \cite{Steinberg=1995ANG}.
It it possible that a portion of this effect
could be due to absorption or scattering in the dielectric
layers at the several percent level,
although further work is necessary to see if such a model can be
tailored to agree with the observed reflection characteristics.

The present results may also be of use in constructing high-reflectivity
dielectric mirrors, whose characteristics are ultimately limited by the (small)
losses in the dielectrics.  It is intuitively clear that the reflectivity is
sensitive
mostly to losses near the entrance face of the mirror, but one might have
expected the transmissivity to be equally sensitive to losses anywhere in
the structure.  The present results show instead that losses near the middle of
the
mirror can be tolerated without having a significant impact on the
transmission,
aside from introducing an additional time delay.  (It is interesting to
note that gain, on the other hand, could shorten the time delay without
having a significant effect on the transmission probability; cf.
\cite{Chiao=1993SUP,Bolda=1994PROP,Steinberg=1994DOUB,Chiao=1994QO}.)
If a ``smooth'' barrier were
constructed, so that the WKB approximation held, the conditional dwell time
would become pure imaginary between the classical turning points, and the
transmission would (in the limit) become completely insensitive to loss in this
region.

In conclusion, we have shown that a straightforward definition of
conditional probabilities in quantum theory, equivalent to Aharonov
{\it et al.}'s idea of ``weak measurements,'' allows one to discuss
the history of a particle which has tunnelled.  This approach makes
clear the meaning of imaginary and complex times, and can describe
the results of a broad range of hypothetical experiments.  It is found
that tunneling particles spend equal amounts of time near the entrance
and exit faces of the barrier, but vanishingly little in the center;
by contrast, reflected particles spend most of their time near the
entrance face only.  One ``clock'' considered is that of absorption
in some region within the barrier.
The anomalously small dwell time is shown to lead to anomalously small
losses, while the imaginary part of the ``conditional'' dwell time
may lead to an additional time delay in the presence of absorption.

	This work was supported by the U.S. Office of Naval Research under
grant N00014-90-J-1259.  I would like to thank R. Y. Chiao for years of
discussions without which this work would have been either impossible
or completed much sooner.  I would also like to acknowledge many useful
comments made by R. Landauer, M. Mitchell, G. Kurizki and Y. Japha.

\newpage
{\bf Figure Captions}
\vspace{0.5in}

1.  A {\it gedankenexperiment} using distant electrons to
measure how much time a
 tunneling proton spends in each of several shielded regions of space.  The
proton is tunneling along $x$, through a series of plates held at some
repulsive
potential $V$.  While the proton
 is between a given pair of plates, only the electron
between the same pair of plates feels a significant Coulomb attraction.
After the proton has stopped interacting with the barrier region, the momentum
of the electrons serves as a record of how much time the proton has spent in
each sub-region.  If we examine the electrons only after detecting a reflected
proton or only after detecting a transmitted proton, we may observe the
time-integrated conditional probability distributions discussed in the text.
(If
time-dependent shutters were added before the electrons, the distributions
could even be measured as a function of time.)
In (a)
we see the final state of the electrons for cases where the proton is
reflected:
only those in the first evanescent decay length of the tunnel barrier feel the
 proton's Coulomb potential and acquire a consequent momentum kick,
indicated by arrows.  In (b)
 we see what happens if the proton is transmitted: electrons near both edges of
the barrier acquire a momentum kick, but those near the center do not.  The
 tunneling proton seems not to have spent any time in the center of the
barrier.
  However, the {\it position} of the electrons gets shifted when the proton is
 transmitted, as indicated by the redrawn electron wave packets.
As explained in the text, this is a indicative of the back-action
of the measuring electrons on the tunneling particle.

2.  The heavy solid curves show the real part of the ``conditional
probability''
for  a tunneling particle to be at position $x$, for several different times.
This represents the force which an electron
at a given x would experience in the example of Figure 1.  The dashed
curve represents the back-action, i.e., the position shift an electron would
 experience.  For comparison, the light solid curve shows the probability
 distribution for the ensemble as a whole, without distinguishing between
 transmitted and reflected particles.  With $\hbar=m=1$, the parameters for
these curves are incident wave vector (and hence initial velocity) $k=0.5$,
barrier height in wave vector units $k_0=0.75$, rms wave vector
uncertainty $\Delta k = 0.03$, and barrier thickness $d=10$ (centered at
$x=0$).
The transmission probability is thus $5.5\cdot 10^{-5}$, and the group
delay for both transmission and reflection is $7.1$, making the effective
traversal speed about three times the free propagation velocity.  For thick
enough barriers, this effective velocity could even exceed the speed of light.
Note that the real part of the
conditional probability distribution mimics the incident wave
packet at early times and the transmitted packet at late times, thus travelling
anomalously fast as well; it appears to do this by ``skipping'' the region in
the middle of the barrier entirely (although while tunneling, a large
 {\it imaginary} part builds up in that region).  Despite the fact that for
certain
parameters, the emerging peak may only be causally connected to the leading
edge of the incident packet, the ({\it measurable}) conditional probabilities
as defined here do not support the identification of the transmitted particles
with the the leading edge of the incident packet.

3.  Same as Figure 2, but for allowed transmission.  In addition, the light
dashed and dotted curves show the conditional probability distribution for
reflected particles.  The parameters are $k=0.75$, $k_0=0.5$,
$\Delta k = 0.02$, and $d=10$, for a reflection probability of $0.035$
and a group delay of $18.2$.  In this case, aside from self-interference
terms, the transmitted-particle distribution traverses the barrier relatively
smoothly, while the reflected-particle distribution displays rapid oscillations
in and to both sides of the barrier region, only reconstructing a smooth
reflected peak when the wave packet has left the barrier region.

4. Numerical calculation of the effect
 of introducing a 5\% amplitude loss into one of the
11 dielectric layers of the mirror studied in \cite{Steinberg=1993PRL}.
As explained in the text, the consequent attenuation of the transmitted and
reflected beams may be thought of as a measure of the time spent in the
layer in question by transmitted and reflected particles, respectively.
The logarithm of this attenuation is plotted against the position of the
lossy layer.  In (a), the incident light is near the center of the bandgap.
In (b), it is near the first resonant transmission point, $k/k_0 \approx 1.2$.
In (c), it is still in a regime of allowed transmission, but with
Fabry-Perot-like
interference lowering the transmission closer to 66\%.  Here $k/k_0 \approx
 1.3$.

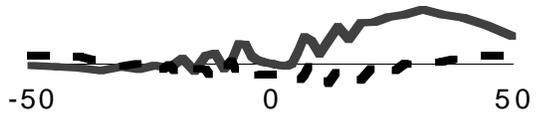
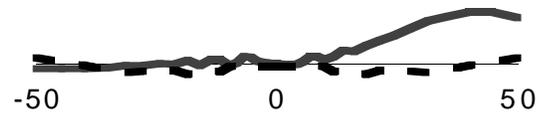
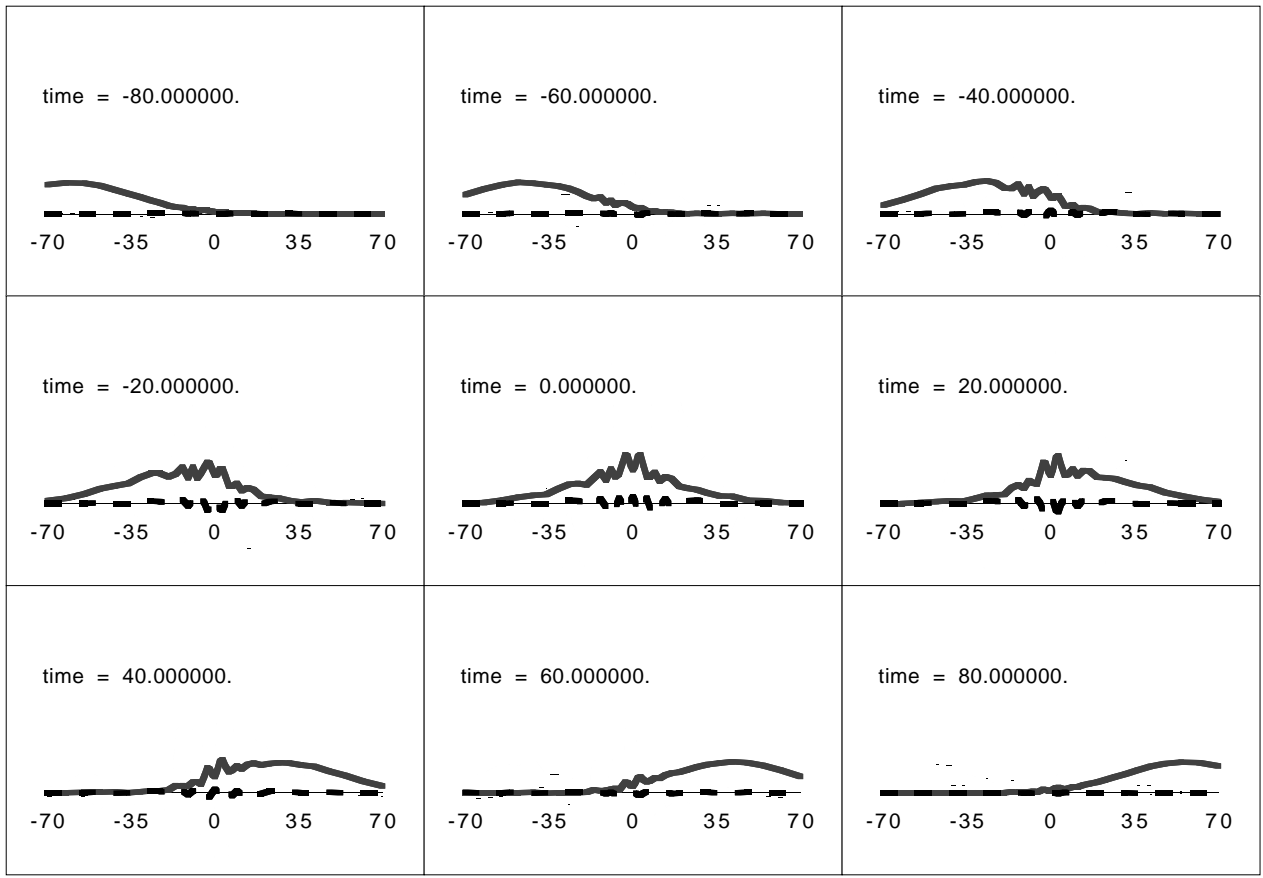
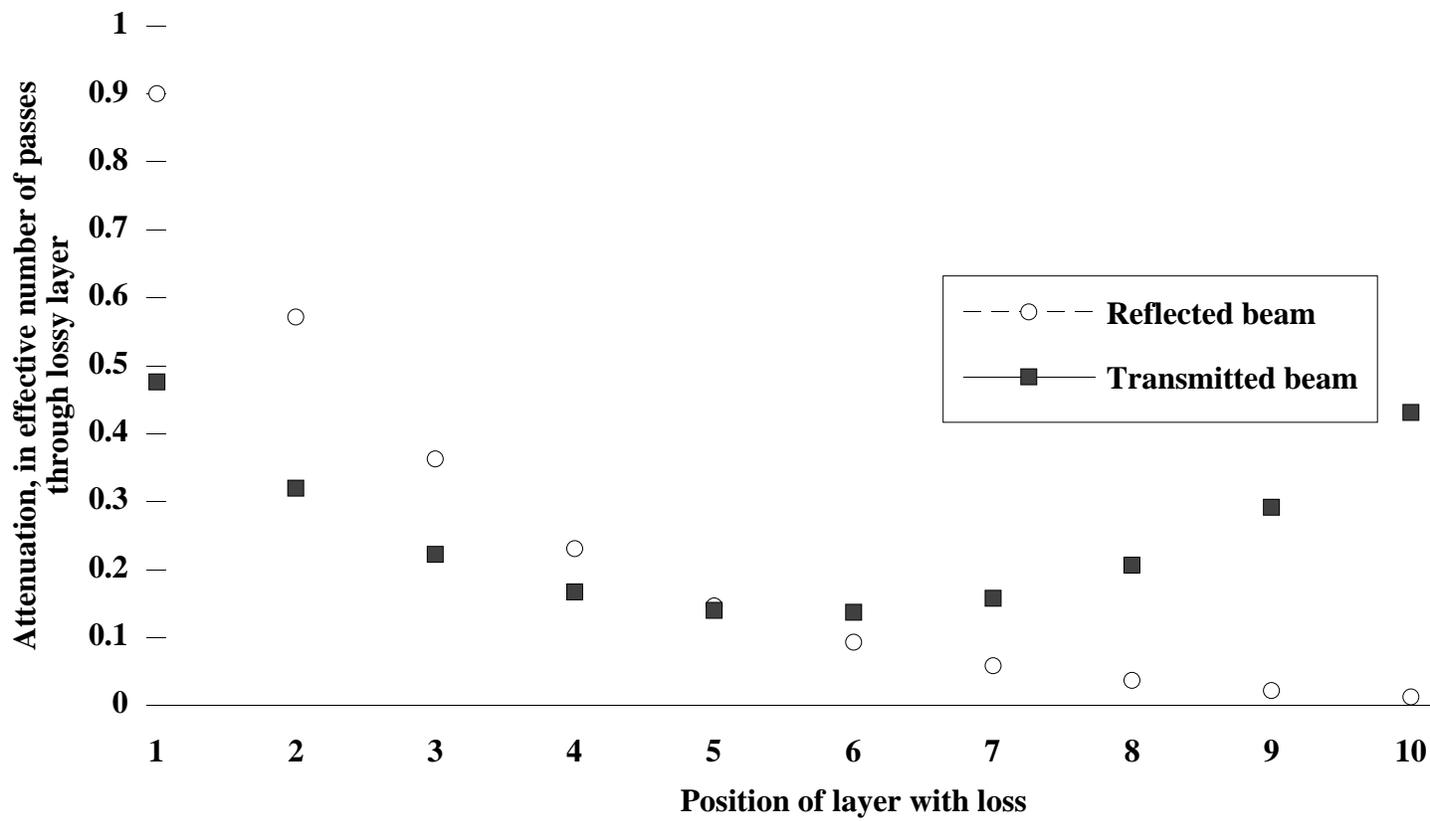
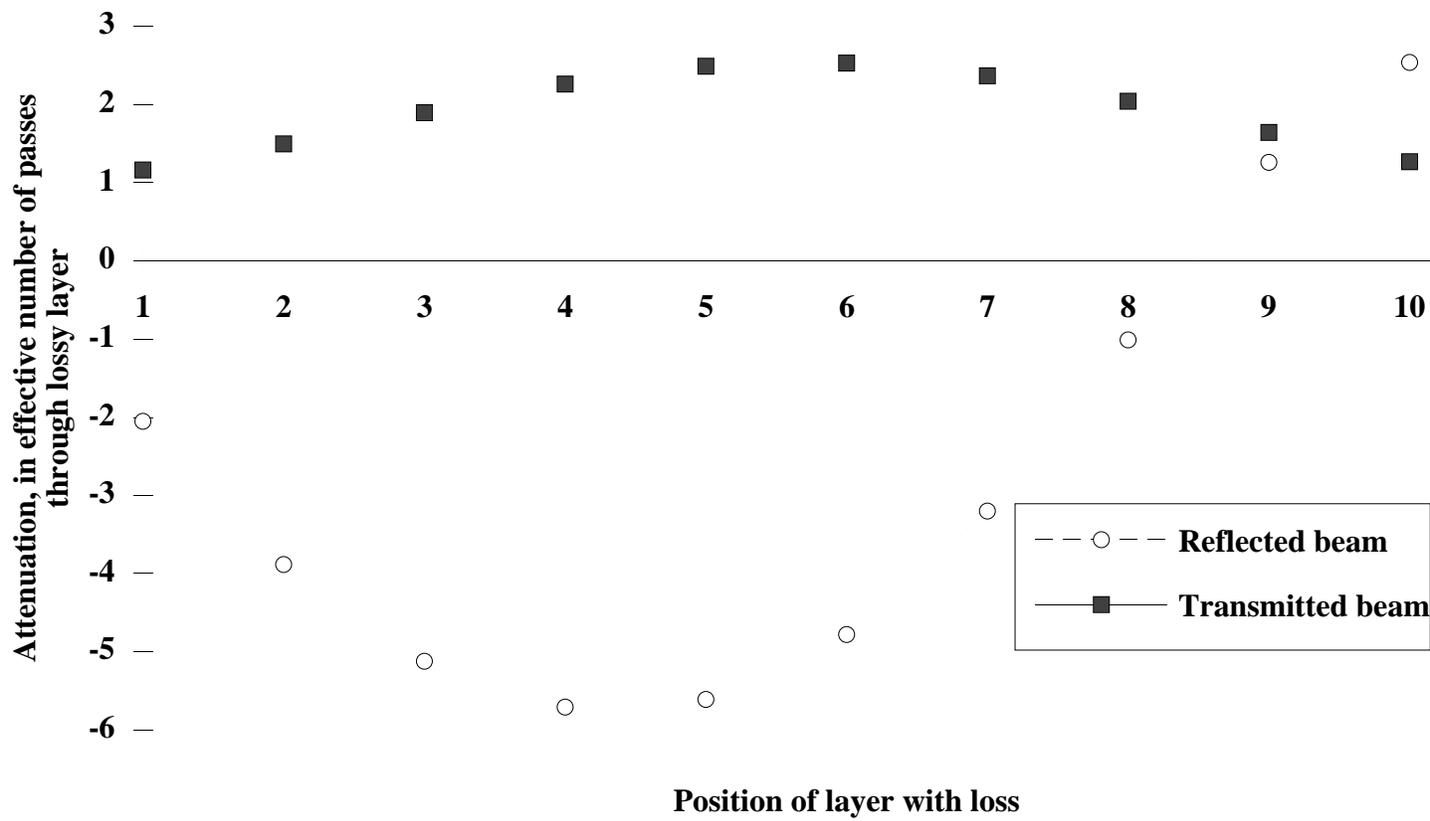
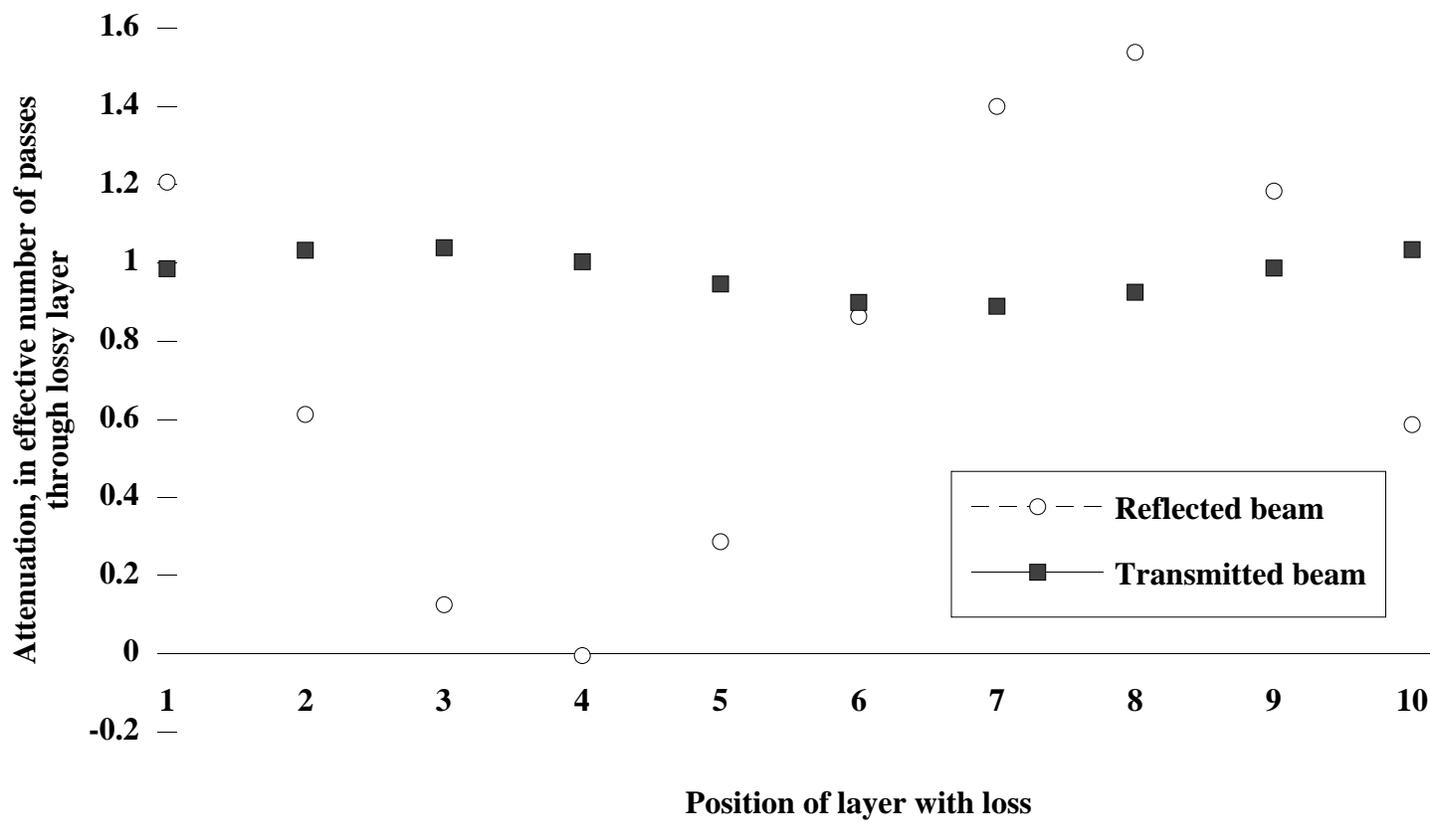
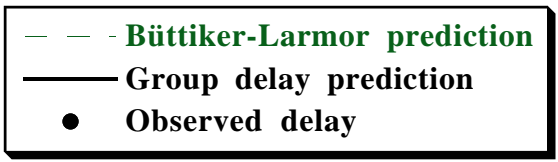
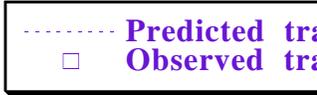
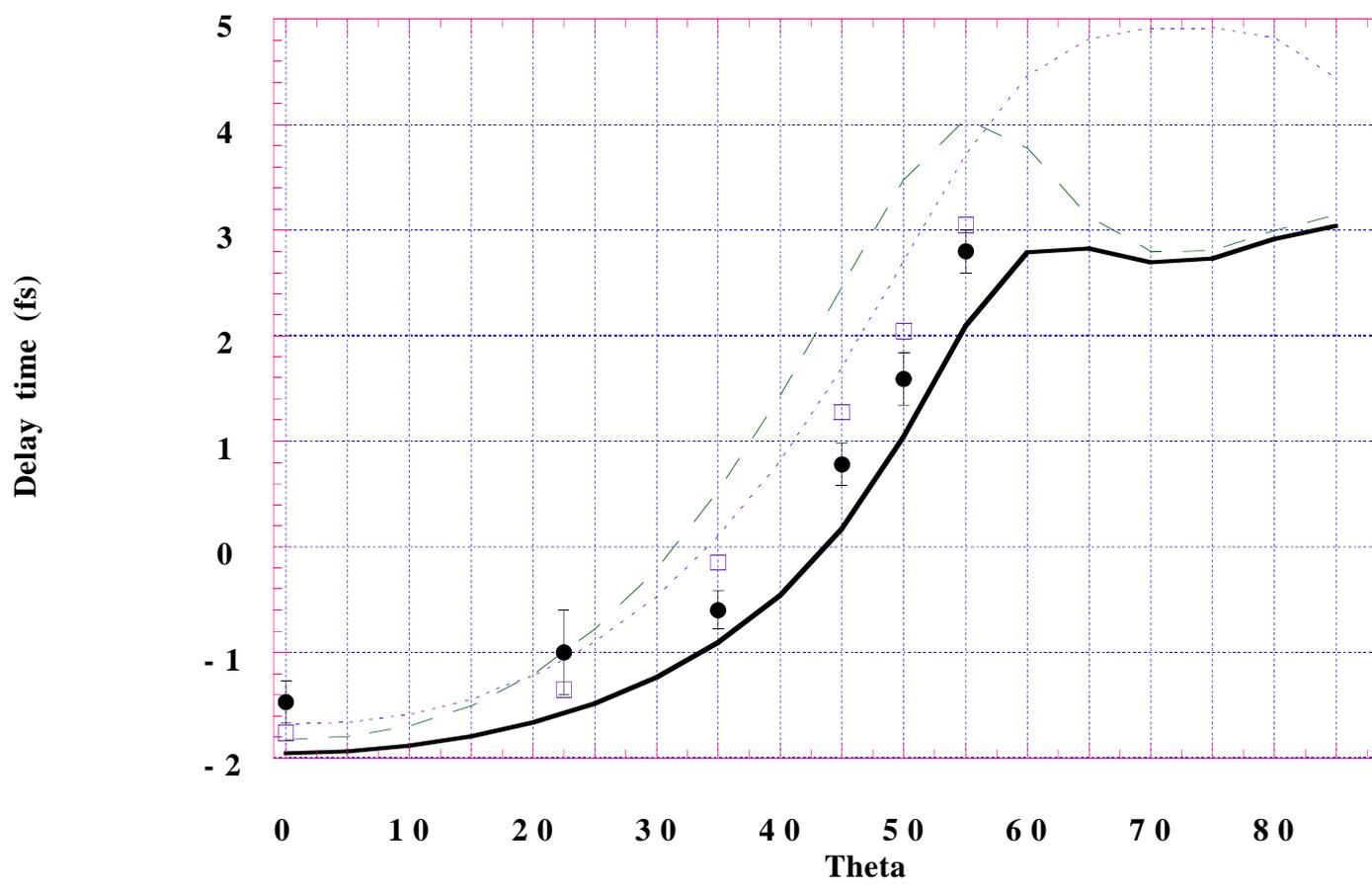
\end{document}